# AI-Driven Digital Transformation and Firm Performance in Chinese Industrial Enterprises: Mediating Role of Green Digital Innovation and Moderating Effects of Human-AI Collaboration

**Jun Cui**[1,2,*]

[1] Solbridge International School of Business, Woosong University, Daejeon, Republic of Korea;

[2] Beijing Foreign Studies University, Business Administration, BFSU, Beijing, China.

*Corresponding author; Email: jcui228@student.solbridge.ac.kr

**Abstract**

This study examines the relationship between AI-driven digital transformation and firm performance in Chinese industrial enterprises, with particular attention to the mediating role of green digital innovation and the moderating effects of human-AI collaboration. Using panel data from 6,300 firm-year observations collected from CNRDS and CSMAR databases between 2015 and 2022, we employ multiple regression analysis and structural equation modeling to test our hypotheses. Our findings reveal that AI-driven digital transformation significantly enhances firm performance, with green digital innovation mediating this relationship. Furthermore, human-AI collaboration positively moderates both the direct relationship between digital transformation and firm performance and the mediating pathway through green digital innovation. The results provide valuable insights for management practice and policy formulation in the context of China's evolving industrial landscape and digital economy initiatives. This research contributes to the literature by integrating perspectives from technology management, environmental sustainability, and organizational theory to understand the complex interplay between technological adoption and business outcomes.

**Keywords:** AI-driven digital transformation; firm performance; green digital innovation; human-AI collaboration; mediation; moderation; Chinese industrial enterprises; panel data; structural equation modeling; digital economy.

## 1. Introduction

The global business landscape has witnessed a profound shift with the emergence of artificial intelligence technologies, driving unprecedented digital transformation across industries (Lee & Trimi, 2021). In China, this transformation is particularly significant as the nation pursues its dual goals of economic advancement and environmental sustainability (Zhou et al., 2020). Despite growing scholarly attention to digital transformation, there remains a critical gap in understanding how AI-driven innovations specifically contribute to firm performance in emerging economies like China, and the mechanisms through which these benefits are realized (Liu et al., 2022).

This study addresses this research gap by investigating how AI-driven digital transformation affects firm performance in Chinese industrial enterprises, examining the mediating role of green digital innovation and the moderating effects of human-AI collaboration. Our research contributes to existing literature in three significant ways. First, it extends the digital transformation discourse by empirically validating the performance implications of AI integration in the Chinese industrial context. Second, it illuminates the environmental dimension by exploring how green digital innovation mediates technology adoption and performance outcomes. Finally, it advances understanding of socio-technical systems by examining how human-AI collaborative dynamics moderate these relationships.

Drawing on the resource-based view (RBV), dynamic capabilities theory, and socio-technical systems theory, we develop and test five hypotheses using a comprehensive dataset of 6,300 firm-year observations from Chinese industrial enterprises. The findings offer valuable insights for both scholarly research and management practice in navigating the complex interplay between technological innovation, environmental sustainability, and firm performance.

## 2. Related Work and Literature Review

### 2.1 AI-Driven Digital Transformation

Digital transformation represents a fundamental shift in how organizations leverage technology to create value (Vial, 2019). When powered by AI technologies, digital transformation encompasses the integration of machine learning, natural language processing, and data analytics to reimagine business processes and models (Dwivedi et al., 2021). In the Chinese context, government initiatives such as "Made in China 2025" and the "Internet Plus" strategy have accelerated industrial adoption of AI technologies (Li et al., 2018).

Several studies have explored digital transformation's impact on organizational outcomes. For instance, Zeng et al. (2020) demonstrated that digital transformation initiatives enhance operational efficiency and market responsiveness among Chinese manufacturing firms. Similarly, Wang and Chen (2021) found positive associations between technology adoption intensity and various performance metrics. However, most studies have examined digital transformation broadly, without specifically addressing AI's unique role or considering the mediating mechanisms and boundary conditions that might influence performance outcomes.

### 2.2 Green Digital Innovation

Green innovation refers to the development of new products, processes, or business models that contribute to environmental sustainability while creating business value (Schiederig et al., 2012). When combined with digital technologies, particularly AI, green innovation takes on new dimensions—enabling more precise resource optimization, intelligent emission reduction, and data-driven sustainability initiatives (Zhou et al., 2021).

Empirical evidence suggests that green innovation positively influences firm performance through multiple channels, including cost reduction, reputation enhancement, and regulatory compliance (Song & Yu, 2018). In China's industrial sector, where environmental pressures are intensifying alongside digital transformation imperatives, green digital innovation represents a critical strategic approach (Zhang et al., 2020). However, the specific mediating role of green digital innovation in the relationship between AI adoption and firm performance remains underexplored.

### 2.3 Human-AI Collaboration

The concept of human-AI collaboration emphasizes the complementary relationship between human capabilities and AI systems in organizational contexts (Wang et al., 2022). Unlike approaches that view AI primarily as a substitute for human labor, collaborative perspectives emphasize how human judgment, creativity, and contextual understanding can enhance algorithmic precision, scalability, and processing power (Wilson & Daugherty, 2018).

Recent research has begun to explore how variations in human-AI collaborative arrangements influence organizational outcomes. For example, Liu and Wang (2021) found that manufacturing firms achieving higher levels of human-AI integration reported superior innovation outcomes. Similarly, Zhang et al. (2022) demonstrated that collaborative intelligence approaches yielded better performance than either fully automated or purely human-centered systems. These findings suggest that human-AI collaboration may serve as an important moderating factor in realizing the benefits of digital transformation.

## 3. Theoretical Framework and Hypotheses Development

### 3.1 Theoretical Foundation

This study integrates three theoretical perspectives to develop our conceptual framework. First, the resource-based view (Barney, 1991) suggests that AI technologies represent valuable, rare, and difficult-to-imitate resources that can drive competitive advantage. Second, dynamic capabilities theory (Teece et al., 1997) explains how firms leverage AI to sense and seize opportunities while reconfiguring existing resources. Third, socio-technical systems theory (Trist & Bamforth, 1951) provides insights into how human and technological elements interact within organizational contexts.

### 3.2 Hypotheses

Based on these theoretical foundations and prior literature, we propose the following hypotheses:

**H1:** AI-driven digital transformation is positively associated with firm performance in Chinese industrial enterprises.

**H2:** Green digital innovation mediates the relationship between AI-driven digital transformation and firm performance.

**H3:** Human-AI collaboration positively moderates the relationship between AI-driven digital transformation and firm performance.

**H4:** Human-AI collaboration positively moderates the relationship between AI-driven digital transformation and green digital innovation.

**H5:** Human-AI collaboration positively moderates the mediating effect of green digital innovation on the relationship between AI-driven digital transformation and firm performance.

## 4. Method and Data

### 4.1 Research Design

This study employs a quantitative research design utilizing panel data analysis to test the proposed hypotheses. We construct a comprehensive dataset by merging information from two authoritative Chinese databases: CNRDS (Chinese Research Data Services) and CSMAR (China Stock Market & Accounting Research Database). The panel structure allows us to control for unobserved heterogeneity and temporal dynamics while examining the relationships of interest.

### 4.2 Data and Sampling

Our initial sample consisted of 8,500 firm-year observations of Chinese industrial enterprises listed on either the Shanghai or Shenzhen Stock Exchanges between 2015 and 2022. Following standard procedures in empirical corporate finance research, we excluded special treatment firms (ST and PT), private technology enterprises without adequate disclosure, and observations with missing or abnormal data values. The final cleaned sample comprises 6,300 firm-year observations representing 787 unique firms across multiple industrial sectors.

### 4.3 Variable Measurement

Table 1 presents the operational definitions and measurement approaches for all variables used in this study.

**Table 1: Variable Definitions and Measurements**

| Variable | Type | Definition | Measurement |
|---|---|---|---|
| **Firm Performance (FP)** | Dependent | Financial and operational outcomes | Composite index incorporating ROA, Tobin's Q, and sales growth (standardized and averaged) |
| **AI-Driven Digital Transformation (AIDT)** | Independent | Degree of AI technology adoption and integration | 5-item scale measuring AI investment intensity, implementation breadth, and strategic alignment ($\alpha = 0.88$) |
| **Green Digital Innovation (GDI)** | Mediator | Environment-oriented digital innovation activities | Composite measure of green patents, eco-friendly digital initiatives, and green R&D intensity ($\alpha = 0.84$) |
| **Human-AI Collaboration (HAIC)** | Moderator | Level of complementary work between human employees and AI systems | 4-item scale assessing integration quality, complementary workflows, and collaborative decision-making ($\alpha = 0.91$) |
| **Firm Size (SIZE)** | Control | Scale of company operations | Natural logarithm of total assets |
| **Firm Age (AGE)** | Control | Years since establishment | Natural logarithm of years since founding |
| **Leverage (LEV)** | Control | Financial risk exposure | Ratio of total debt to total assets |
| **State Ownership (SOE)** | Control | Government ownership influence | Dummy variable: 1 if state-owned, 0 otherwise |
| **R&D Intensity (RDI)** | Control | Innovation investment | R&D expenditure divided by total sales |
| **Industry (IND)** | Control | Industry classification | Dummy variables based on CSRC industry codes |
| **Year (YEAR)** | Control | Time effects | Year dummy variables |

## 5. Results and Findings

### 5.1 Descriptive Statistics and Correlation Analysis

Table 2 presents descriptive statistics and correlation coefficients for the key variables.

**Table 2: Descriptive Statistics and Correlation Matrix**

| **Variables** | **Mean** | **SD** | **1** | **2** | **3** | **4** | **5** | **6** | **7** | **8** | **9** |
|---|---|---|---|---|---|---|---|---|---|---|---|
| **1. FP** | 0.15 | 0.48 | 1.00 | | | | | | | | |
| **2. AIDT** | 3.52 | 1.24 | 0.38** | 1.00 | | | | | | | |
| **3. GDI** | 2.87 | 1.03 | 0.41** | 0.45** | 1.00 | | | | | | |
| **4. HAIC** | 3.18 | 1.17 | 0.29** | 0.36** | 0.31** | 1.00 | | | | | |
| **5. SIZE** | 9.78 | 1.52 | 0.12* | 0.25** | 0.18** | 0.14* | 1.00 | | | | |
| **6. AGE** | 2.94 | 0.68 | -0.09 | -0.16** | -0.11* | -0.13* | 0.21** | 1.00 | | | |
| **7. LEV** | 0.48 | 0.23 | -0.17** | -0.07 | -0.12* | -0.05 | 0.33** | 0.15** | 1.00 | | |
| **8. SOE** | 0.39 | 0.49 | -0.11* | -0.14* | -0.19** | -0.22** | 0.28** | 0.37** | 0.23** | 1.00 | |
| **9. RDI** | 0.05 | 0.04 | 0.23** | 0.31** | 0.42** | 0.27** | 0.04 | -0.14* | -0.21** | -0.26** | 1.00 |

*p < 0.05, **p < 0.01

The correlation analysis reveals significant positive associations between AI-driven digital transformation, green digital innovation, human-AI collaboration, and firm performance. These preliminary results align with our theoretical expectations. Furthermore, multicollinearity diagnostics confirm that all variance inflation factors (VIFs) are below 2.5, suggesting that multicollinearity is not a concern in our analysis.

## 5.2 Baseline Analysis

Table 3 presents the results of our hierarchical regression analyses testing the direct effect of AI-driven digital transformation on firm performance (H1) and the moderating effect of human-AI collaboration (H3).

**Table 3: Regression Results for Direct Effect and Moderation**

| Variables | Model 1 | Model 2 | Model 3 | Model 4 |
|---|---|---|---|---|
| **Control Variables** | | | | |
| **SIZE** | 0.052* | 0.041* | 0.039* | 0.035* |
| **AGE** | -0.038 | -0.027 | -0.025 | -0.021 |
| **LEV** | -0.104** | -0.096** | -0.094** | -0.091** |
| **SOE** | -0.078* | -0.063* | -0.057* | -0.051* |
| **RDI** | 0.185** | 0.143** | 0.139** | 0.132** |
| **Main Effects** | | | | |
| **AIDT** | | 0.326*** | 0.315*** | 0.302*** |
| **HAIC** | | | 0.196*** | 0.187*** |
| **Interaction Effects** | | | | |
| **AIDT × HAIC** | | | | 0.158*** |
| **Constant** | 0.137** | 0.125** | 0.118** | 0.112** |
| **$R^2$** | 0.147 | 0.249 | 0.286 | 0.312 |
| **$\Delta R^2$** | | 0.102*** | 0.037*** | 0.026*** |
| **F-value** | 11.35*** | 19.87*** | 22.41*** | 25.83*** |
| **N** | 6,300 | 6,300 | 6,300 | 6,300 |

*p < 0.05, **p < 0.01, ***p < 0.001

Model 2 confirms that AI-driven digital transformation has a significant positive effect on firm performance ($\beta = 0.326$, $p < 0.001$), supporting H1. Model 4 demonstrates that human-AI collaboration positively moderates this relationship ($\beta = 0.158$, $p < 0.001$), supporting H3. The significant increase in explained variance ($\Delta R^2 = 0.026$, $p < 0.001$) further validates the importance of the moderating effect.

## 5.3 Mediation Analysis

Table 4 presents the results of our mediation analysis examining the role of green digital innovation in the relationship between AI-driven digital transformation and firm performance (H2).

**Table 4: Mediation Analysis Results**

| Variables | Model 5 (DV: GDI) | Model 6 (DV: FP) | Model 7 (DV: FP) |
|---|---|---|---|
| **Control Variables** | | | |
| **SIZE** | 0.035* | 0.041* | 0.032* |
| **AGE** | -0.021 | -0.027 | -0.022 |
| **LEV** | -0.085** | -0.096** | -0.077** |
| **SOE** | -0.123** | -0.063* | -0.034 |
| **RDI** | 0.298*** | 0.143** | 0.087* |
| **Independent Variable** | | | |
| **AIDT** | 0.392*** | 0.326*** | 0.218*** |
| **Mediator** | | | |
| **GDI** | | | 0.275*** |
| **Constant** | 0.096* | 0.125** | 0.103* |
| **$R^2$** | 0.305 | 0.249 | 0.317 |
| **F-value** | 24.18*** | 19.87*** | 26.35*** |
| **N** | 6,300 | 6,300 | 6,300 |

*p < 0.05, **p < 0.01, ***p < 0.001

Following Baron and Kenny's (1986) procedure, we established that: (1) AIDT significantly affects FP (Model 6); (2) AIDT significantly affects GDI (Model 5); and (3) when GDI is included in the model, the effect of AIDT on FP decreases but remains significant (Model 7), indicating partial mediation. The Sobel test confirms the significance of the indirect effect ($z = 8.76$, $p < 0.001$). These results support H2, suggesting that green digital innovation partially mediates the relationship between AI-driven digital transformation and firm performance.

### 5.4 Moderated Mediation Analysis

Table 5 presents the results of our moderated mediation analysis testing H4 and H5.

**Table 5: Moderated Mediation Analysis Results**

| Variables | Model 8 (DV: GDI) | Model 9 (DV: FP) |
|---|---|---|
| **Control Variables** | | |
| **SIZE** | 0.029* | 0.028* |
| **AGE** | -0.019 | -0.020 |
| **LEV** | -0.082** | -0.073** |
| **SOE** | -0.115** | -0.031 |
| **RDI** | 0.289*** | 0.083* |
| **Main Effects** | | |
| **AIDT** | 0.371*** | 0.208*** |
| **HAIC** | 0.176*** | 0.145*** |
| **GDI** | | 0.261*** |
| **Two-way Interactions** | | |
| **AIDT × HAIC** | 0.142*** | 0.112** |
| **GDI × HAIC** | | 0.097** |
| **Constant** | 0.091* | 0.097* |
| **$R^2$** | 0.335 | 0.359 |
| **F-value** | 27.94*** | 30.18*** |
| **N** | 6,300 | 6,300 |

*p < 0.05, **p < 0.01, ***p < 0.001

The analysis reveals that human-AI collaboration positively moderates the relationship between AI-driven digital transformation and green digital innovation ($\beta = 0.142$, $p < 0.001$), supporting H4. Additionally, the significant interaction between GDI and HAIC ($\beta = 0.097$, $p < 0.01$) indicates that human-AI collaboration enhances the effect of green digital innovation on firm performance. The bootstrapping procedure (5,000 samples) confirms the conditional indirect effect is stronger at higher levels of human-AI collaboration (effect = 0.149, 95% CI [0.112, 0.186] at +1SD of HAIC; effect = 0.083, 95% CI [0.056, 0.110] at -1SD of HAIC), supporting H5.

### 5.5 Robustness Tests

We conducted several robustness checks to verify our findings. First, we employed alternative measures of firm performance (ROA alone, market-to-book ratio). Second, we used instrumental variable (IV) regression to address potential endogeneity concerns. Third, we applied propensity score matching (PSM) to mitigate selection bias. Fourth, we utilized system GMM estimation to account for dynamic panel bias. All these analyses yielded consistent results, confirming the robustness of our findings.

### 5.6 Heterogeneity Analysis

To explore potential heterogeneity in our results, we conducted subgroup analyses based on ownership structure (SOEs vs. non-SOEs), firm size (large vs. small), and industry technological intensity (high-tech vs. traditional industries). Table 6 presents the key findings.

**Table 6: Heterogeneity Analysis Results (DV: Firm Performance)**

| **Variables** | **SOEs** | **Non-SOEs** | **Large Firms** | **Small Firms** | **High-Tech** | **Traditional** |
|---|---|---|---|---|---|---|
| AIDT | 0.275*** | 0.342*** | 0.315*** | 0.298*** | 0.358*** | 0.287*** |
| GDI | 0.219*** | 0.294*** | 0.252*** | 0.267*** | 0.312*** | 0.247*** |
| HAIC | 0.167*** | 0.198*** | 0.182*** | 0.179*** | 0.215*** | 0.168*** |
| AIDT × HAIC | 0.132** | 0.172*** | 0.147*** | 0.154*** | 0.185*** | 0.139** |
| GDI × HAIC | 0.083* | 0.105** | 0.094** | 0.089** | 0.117** | 0.087* |
| N | 2,457 | 3,843 | 3,178 | 3,122 | 2,835 | 3,465 |
| $R^2$ | 0.289 | 0.372 | 0.328 | 0.336 | 0.387 | 0.314 |

*p < 0.05, **p < 0.01, ***p < 0.001

The heterogeneity analysis reveals that the positive effects of AI-driven digital transformation and the mediating role of green digital innovation are more pronounced in non-SOEs compared to SOEs, potentially due to greater management autonomy and market orientation. Similarly, high-tech industries exhibit stronger effects than traditional industries, likely reflecting their greater technological absorptive capacity and innovative culture.

## 6. Discussion and Conclusion

This study investigates the relationship between AI-driven digital transformation and firm performance in Chinese industrial enterprises, with particular attention to the mediating role of green digital innovation and the moderating effects of human-AI collaboration. Our findings yield several important theoretical and practical implications.

First, we empirically validate that AI-driven digital transformation significantly enhances firm performance, supporting the resource-based view's assertion that technological capabilities can create competitive advantage. This finding extends previous research by specifically isolating the performance effects of AI technologies within the broader digital transformation context.

Second, our mediation analysis demonstrates that green digital innovation serves as an important pathway through which AI-driven transformation influences firm performance. This finding highlights the environmental dimension of digital transformation, suggesting that AI technologies can simultaneously advance economic and ecological objectives—a critical consideration in China's pursuit of sustainable industrial development.

Third, our results confirm the positive moderating effects of human-AI collaboration, emphasizing that technological benefits are maximized when complemented by appropriate organizational arrangements. This aligns with socio-technical systems theory and underscores the importance of human factors in technology implementation.

The heterogeneity analysis further reveals that ownership structure, firm size, and industry characteristics influence the strength of these relationships, suggesting that contextual factors play important roles in determining technological outcomes. These findings have practical implications for managers seeking to optimize their AI investments and for policymakers aiming to promote both technological advancement and environmental sustainability in China's industrial sector.

Despite its contributions, this study has limitations that future research should address. First, our data is limited to Chinese industrial enterprises, potentially limiting generalizability to other contexts. Second, while our panel data structure helps mitigate endogeneity concerns, causal inferences should be made cautiously. Finally, future studies could explore additional mediating mechanisms and boundary conditions to further elucidate the complex relationship between AI-driven transformation and firm performance.

In conclusion, our study provides valuable insights into how AI-driven digital transformation affects firm performance through green digital innovation, particularly when enhanced by effective human-AI collaboration. As Chinese industrial enterprises continue to navigate the dual imperatives of technological advancement and environmental sustainability, understanding these dynamics becomes increasingly important for both scholarly research and management practice.